\newcommand\aj{{AJ\,}}%
\newcommand\apj{{ApJ\,}}%
\newcommand\apjl{{ApJ\,}}%
\newcommand\aap{{A\&A\,}}%
\newcommand\mnras{{MNRAS\,}}%
\documentclass[graybox, natbib, footinfo, fleqn]{svmult}

\usepackage{mathptmx}       % selects Times Roman as basic font
\usepackage{helvet}         % selects Helvetica as sans-serif font
\usepackage{courier}        % selects Courier as typewriter font
\usepackage{type1cm}        % activate if the above 3 fonts are
\usepackage{makeidx}         % allows index generation
\usepackage{graphicx}        % standard LaTeX graphics tool
\usepackage{multicol}        % used for the two-column index
\usepackage[bottom]{footmisc}% places footnotes at page bottom

\makeindex             % used for the subject index
\begin{document}

\title*{Exploiting the open clusters in the {\it Kepler} and CoRoT fields}
\author{Karsten Brogaard, Eric Sandquist, Jens Jessen-Hansen, Frank Grundahl, and S\o ren Frandsen}
\institute{Karsten Brogaard \at Stellar Astrophysics Centre, Department of Physics and Astronomy, Aarhus University, Ny Munkegade 120, 8000 Aarhus C, Denmark, \email{kfb@phys.au.dk} 
\and Eric Sandquist \at Department of Astronomy, San Diego State University, San Diego, CA 92182, USA, \email{erics@mintaka.sdsu.edu}
\and Jens Jessen-Hansen \at Stellar Astrophysics Centre, Department of Physics and Astronomy, Aarhus University, Ny Munkegade 120, 8000 Aarhus C, Denmark, \email{jensjh@phys.au.dk}
\and S\o ren Frandsen \at Stellar Astrophysics Centre, Department of Physics and Astronomy, Aarhus University, Ny Munkegade 120, 8000 Aarhus C, Denmark, \email{srf@phys.au.dk} \and Frank Grundahl \at Stellar Astrophysics Centre, Department of Physics and Astronomy, Aarhus University, Ny Munkegade 120, 8000 Aarhus C, Denmark, \email{fgj@phys.au.dk}
}
%\and Name of Second Author \at Name, Address of Institute \email{name@email.address}}
\authorrunning{Brogaard et al.}
\maketitle

\abstract{
The open clusters in the {\it Kepler} and CoRoT fields potentially provide tight constraints for tests of stellar models and observational methods because they allow a combination of complementary methods. We are in the process of identifying and measuring parameters for detached eclipsing binaries (dEBs) in the open clusters in the {\it Kepler} and CoRoT fields. We make use of measurements of dEBs in the clusters to test the accuracy of asteroseismic scaling relations for mass. We are able to provide strong indications that the asteroseismic scaling relations overestimate the stellar mass, but we are not yet able to distinguish between different proposed corrections from the literature. We argue how our ongoing measurements of more dEBs in more clusters, complemented by dEBs in the field, should be able to break the degeneracy. We also briefly describe how we can identify cluster stars that have evolved through non-standard evolution by making use of ensemble asteroseismology.}
 
\section{Introduction}
\label{sec:1}
Open star clusters are often observed to exploit the advantages of the additional information that comes from observing an ensemble of stars with identical ages and similar metallicities. The open clusters in the {\it Kepler} and CoRoT fields extend the prospects for such an approach because they make the identification of detached eclipsing binary stars much easier and allow us to combine classical observational methods with asteroseismology of giant stars. We are in the process of identifying and measuring parameters for dEBs in the open clusters in the {\it Kepler} (NGC6791, NGC6811, NGC6819, and NGC6866) and CoRoT (NGC6633) fields.

Our long term goal is to test and improve stellar models and gain detailed insights into stellar evolution in the clusters by making use of measurements of multiple dEBs in combination with asteroseismology of single giant stars. In this contribution we use the measurements of dEBs in the clusters to test the accuracy of the asteroseismic scaling relations for mass. 

\section{Tests of asteroseismic scaling relations}
\label{sec:2}
% Always give a unique label
% and use \ref{<label>} for cross-references
% and \cite{<label>} for bibliographic references
% use \sectionmark{}
% to alter or adjust the section heading in the running head

The asteroseismic scaling relations for mass and radius,

\begin{eqnarray}
\frac{M}{\mathrm{M}_\odot} \simeq \left(\frac{\nu _{\mathrm{max}}}{\nu _{\mathrm{max,}\odot}}\right)^3 \left(\frac{\Delta \nu}{\Delta \nu _{\odot}}\right)^{-4} \left(\frac{T_{\mathrm{eff}}}{T_{\mathrm{eff,}\odot}}\right)^{3/2}, \nonumber\\
\frac{R}{\mathrm{R}_\odot} \simeq \left(\frac{\nu _{\mathrm{max}}}{\nu _{\mathrm{max,}\odot}}\right) \left(\frac{\Delta \nu}{\Delta \nu _{\odot}}\right)^{-2} \left(\frac{T_{\mathrm{eff}}}{T_{\mathrm{eff,}\odot}}\right)^{1/2}, 
%a \times b = c \nonumber\\
%\vec{a} \cdot \vec{b}=\vec{c}
\label{eq:01}
\end{eqnarray}

(e.g. ref. \citealt{Miglio12}) provide a relatively easy way of measuring the mass and radius of a star showing solar-like oscillations using a light curve from {\it Kepler} or CoRoT. However, these relations are only approximate and their accuracy is not known in great detail. Moreover, several corrections have been suggested in the literature. Some are based on observations and/or model predictions \citep{Miglio12, White11}, while others try to deal with systematics arising due to not fulfilling assumptions in the derivation of the scaling relations \citep{Mosser13}. The accuracy and precision of the scaling relations and their suggested corrections needs verification.

We wish to employ our measurements of eclipsing binaries in the open clusters for such tests. First rough comparisons of masses between the methods were already done for NCG6791 \citep{Brogaard12} and NGC6819 \citep{Sandquist13}. Those results, shown in Table \ref{tab:1}, indicate that masses from the asteroseismic scaling relations are slightly overestimated.

\begin{table}
\caption{Mean mass of giant stars in clusters from different methods.}
\label{tab:1}       % Give a unique label
%
% Follow this input for your own table layout
%
\begin{tabular}{p{2.6cm}p{2.6cm}p{2.6cm}p{3.0cm}}
\hline\noalign{\smallskip}
Cluster & asteroseismic   		& asteroseismic 		& $<M_{\rm{RGB}}/M_\odot>$\\
        & $<M_{\rm{RGB}}/M_\odot>$	& $<M_{\rm{RGB}}/M_\odot>$	& from eclipsing binaries \\
\noalign{\smallskip}\svhline\noalign{\smallskip}
NGC\,6791	&	$1.22\pm0.01^a$	&	$1.23\pm0.02^b$		&	$1.15\pm0.02^c$ \\
NGC\,6819	&	$1.68\pm0.03^a$	&	$1.61\pm0.04^b$		&	$1.55\pm0.06^d$ \\
%$(m-M)v$			& $12.44\pm0.08^a$	\\
%&$M_{\rm{RGB}}/M_\odot$	& $1.55\pm0.06^b$		\\
\noalign{\smallskip}\hline\noalign{\smallskip}
\end{tabular}
\\$^a$ based on asteroseismic grid-modelling measurements by \cite{Basu11}.
\\$^b$ based on asteroseismic scaling relation measurements by \cite{Miglio12}.
\\$^c$ based on measurements of eclipsing binaries by \cite{Brogaard12}.
\\$^d$ based on measurements of eclipsing binaries by \cite{Jeffries13} and \cite{Sandquist13}.
\end{table}

Here we do a more detailed comparison for NGC6819 and then elaborate on how including the additional clusters will provide improved insights.

\begin{figure} %[b]
%\sidecaption
% For example, with the graphicx style use
\includegraphics[scale=.70]{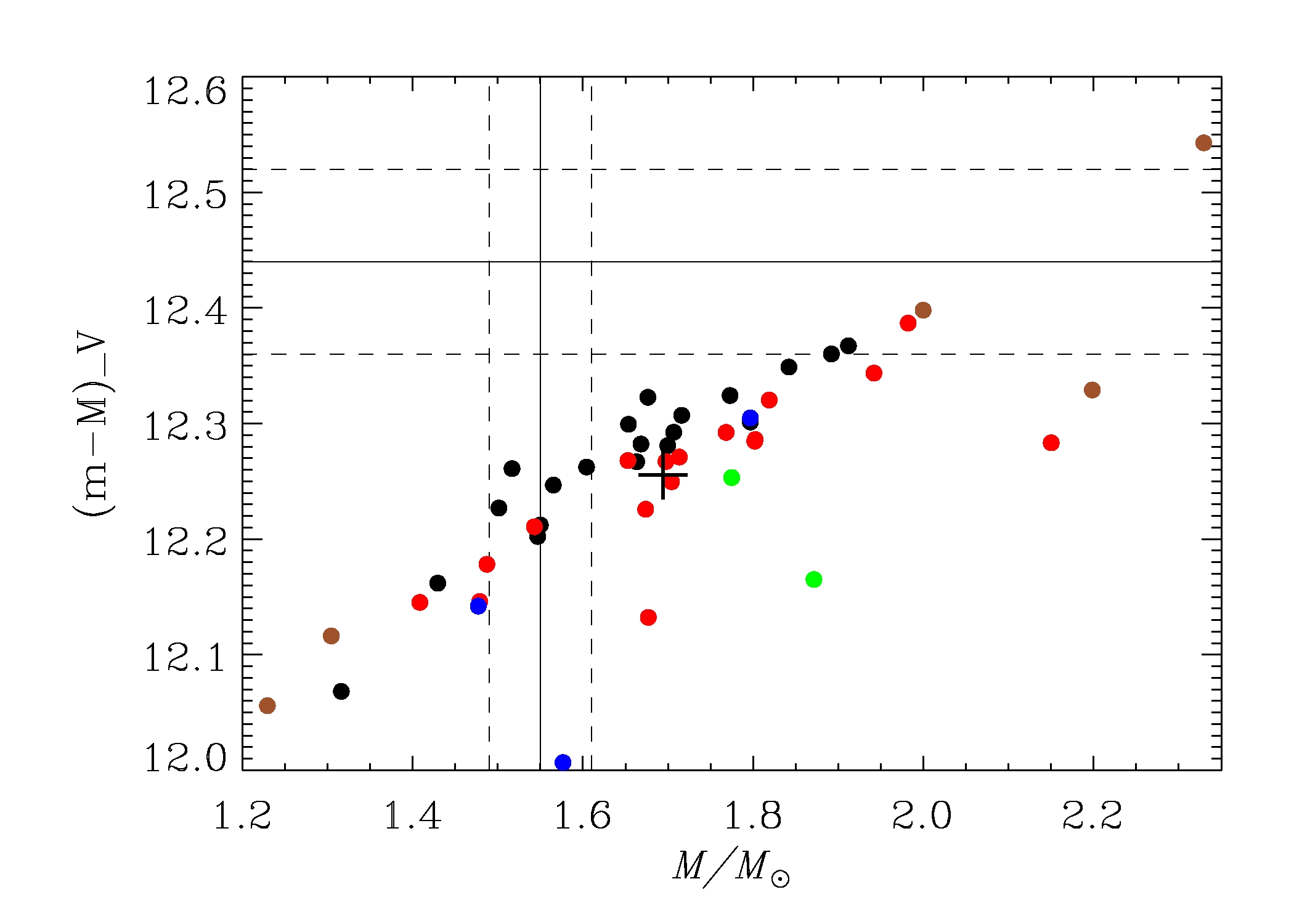}
%
% If no graphics program available, insert a blank space i.e. use
%\picplace{5cm}{2cm} % Give the correct figure height and width in cm
%
\caption{Measurements of mass and distance modulus for giant stars in NGC6819. The solid lines indicate the values inferred from the eclipsing binaries, and the dashed lines the corresponding values $\pm1-\sigma$ uncertainty. Solid dots are values for individual giant stars using the asteroseismic scaling relations in the form of \cite{Miglio12}. Black dots are RGB stars. Brown dots are cool RGB stars with $(V-K_{s})$ colour $\ge$ 3.1, which are close to or above the validity level of the colour-$t_{{\rm eff}}$ relations of \cite{Ramirez05}. Red dots are red clump stars determined from asteroseismology, while blue dots are RC stars determined from the CMD for cases where asteroseismology could not determine the evolutionary phase. Green dots are over-massive RGB stars in binary systems \cite{Corsaro12}. The plus sign marks the mean asteroseismic values, excluding over-massive stars, see Sect. \ref{sec:3}}
\label{fig:1}       % Give a unique label
\end{figure}

In order to get a more detailed view of the situation we show, in Fig. \ref{fig:1}, the measurements in a plot of mass versus apparent distance modulus. The solid lines indicate the measurement of the mass of a star on the red giant branch (RGB) and the cluster distance modulus from the eclipsing binary measurements of \cite{Sandquist13} and \cite{Jeffries13}, slightly adjusted as in Table \ref{tab:2}, due to our use of bolometric corrections from \cite{Casagrande13}. Each circle marks the asteroseismic measurement of mass and distance modulus for a giant star in the cluster, recalculated {\it exactly} as in \cite{Miglio12}, e.g. using the scaling relations in the form of Eq. \ref{eq:01}, with $T_{\rm{eff}}$ values calculated using $V-K_s$ colours and the colour-$T_{\rm{eff}}$ relations of \cite{Ramirez05} and bolometric corrections from \cite{Flower96}. The asteroseismic measurements are from \cite{Corsaro12}.

The plus sign marks the ensemble mean, which is too high compared to the binary measurements. Notice how the asteroseismic measurements fall more or less along a tilted line. This is the consequence of the way the random uncertainties in $T_{\rm{eff}}$ and the seismic parameters affect the position of a star in this diagram; all uncertainties shift the star along approximately the same line. Therefore, there appears to be no way of reaching agreement with the binary result, as the mean asteroseismic values can only shift approximately along the line already defined by the asteroseismic measurements. However, as we shall see, this is an artifact arising due to the use of colour-temperature relations and bolometric corrections from different sources, which underlines the importance of using self-consistent calibrations.

% Use the \index{} command to code your index words

\begin{table}
\caption{NGC6819 parameters determined from eclipsing binary members}
\label{tab:2}       % Give a unique label
%
% Follow this input for your own table layout
%
\begin{tabular}{p{4.5cm}p{4.5cm}}
\hline\noalign{\smallskip}
Parameter & Value \\
\noalign{\smallskip}\svhline\noalign{\smallskip}
$(m-M)_V$			& $12.44\pm0.08^a$	\\
$M_{\rm{RGB}}/M_\odot$	& $1.55\pm0.06^b$		\\
\noalign{\smallskip}\hline\noalign{\smallskip}
\end{tabular}
\\$^a$ recalculated using radii, spectroscopic $T_{\rm{eff}}$s and photometric data for the \\eclipsing binaries from \cite{Jeffries13} and \cite{Sandquist13} with bolometric corrections from \cite{Casagrande13}
\\$^b$ from range of numbers in \cite{Sandquist13}
\end{table}

%\begin{tabular}{p{2cm}p{2.4cm}p{2cm}p{4.9cm}}
%\hline\noalign{\smallskip}
%Classes & Subclass & Length & Action Mechanism  \\
%\noalign{\smallskip}\svhline\noalign{\smallskip}
%Translation & mRNA$^a$  & 22 (19--25) & Translation repression, mRNA cleavage\\
%Translation & mRNA cleavage & 21 & mRNA cleavage\\
%Translation & mRNA  & 21--22 & mRNA cleavage\\
%Translation & mRNA  & 24--26 & Histone and DNA Modification\\
%\end{tabular}
%\end{table}

\begin{figure} %[t]
%\sidecaption
% For example, with the graphicx style use
\includegraphics[scale=.70]{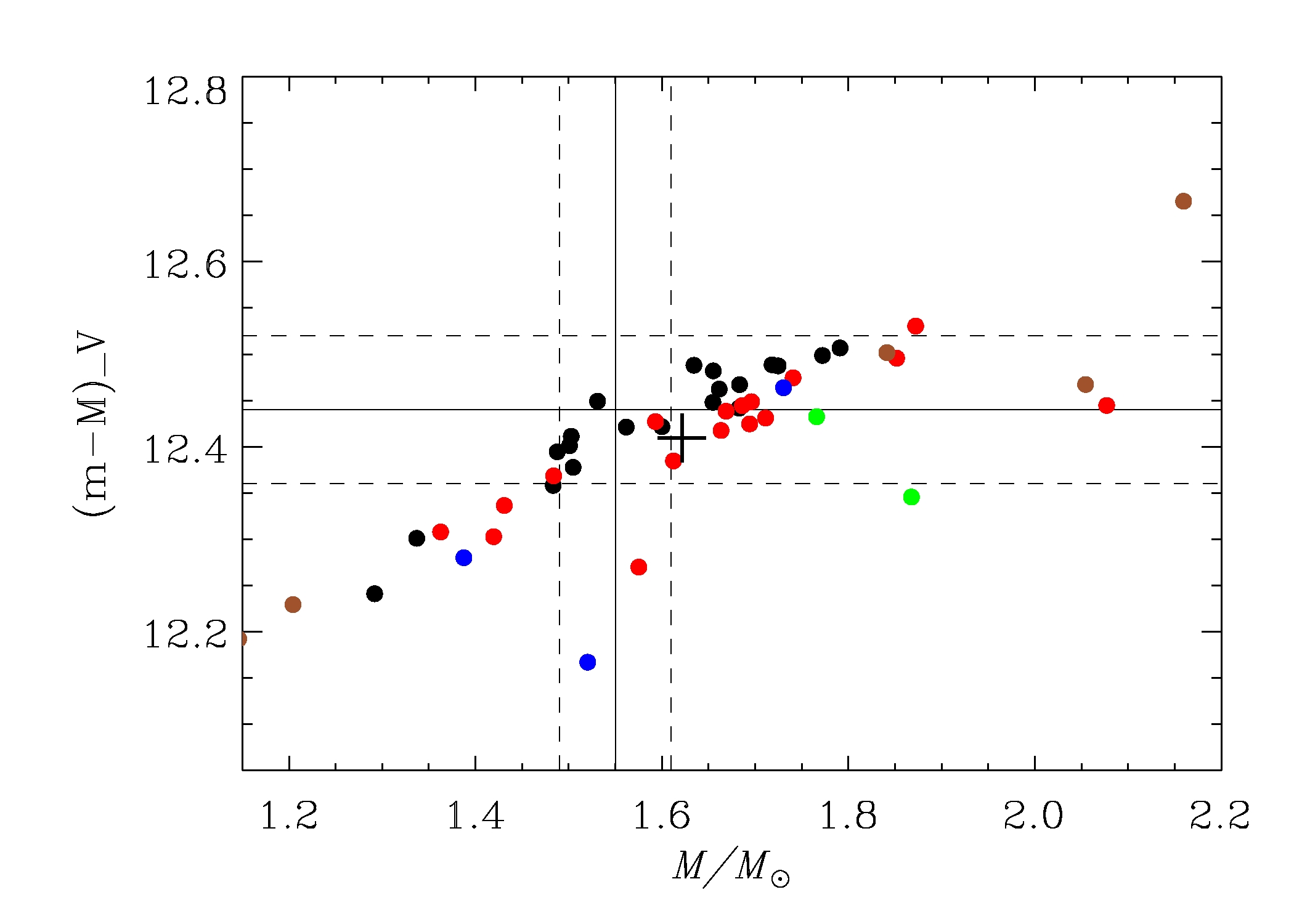}
%
% If no graphics program available, insert a blank space i.e. use
%\picplace{5cm}{2cm} % Give the correct figure height and width in cm
%
\caption{As Fig. \ref{fig:1} but using bolometric corrections and colour-temperature relations from \cite{Casagrande13} and the correction to the scaling relations from \cite{White11}. We use a modified version of their suggested correction, since it is clear from their Fig. 5 that the form of their correction breaks down below $\sim 4700$ K. Below that value we keep the correction constant. The relative correction between RGB and RC stars determined by \cite{Miglio12} is still applied to the RC stars.}
\label{fig:2}       % Give a unique label
\end{figure}

\begin{figure} %[b]
%\sidecaption
% For example, with the graphicx style use
\includegraphics[scale=.70]{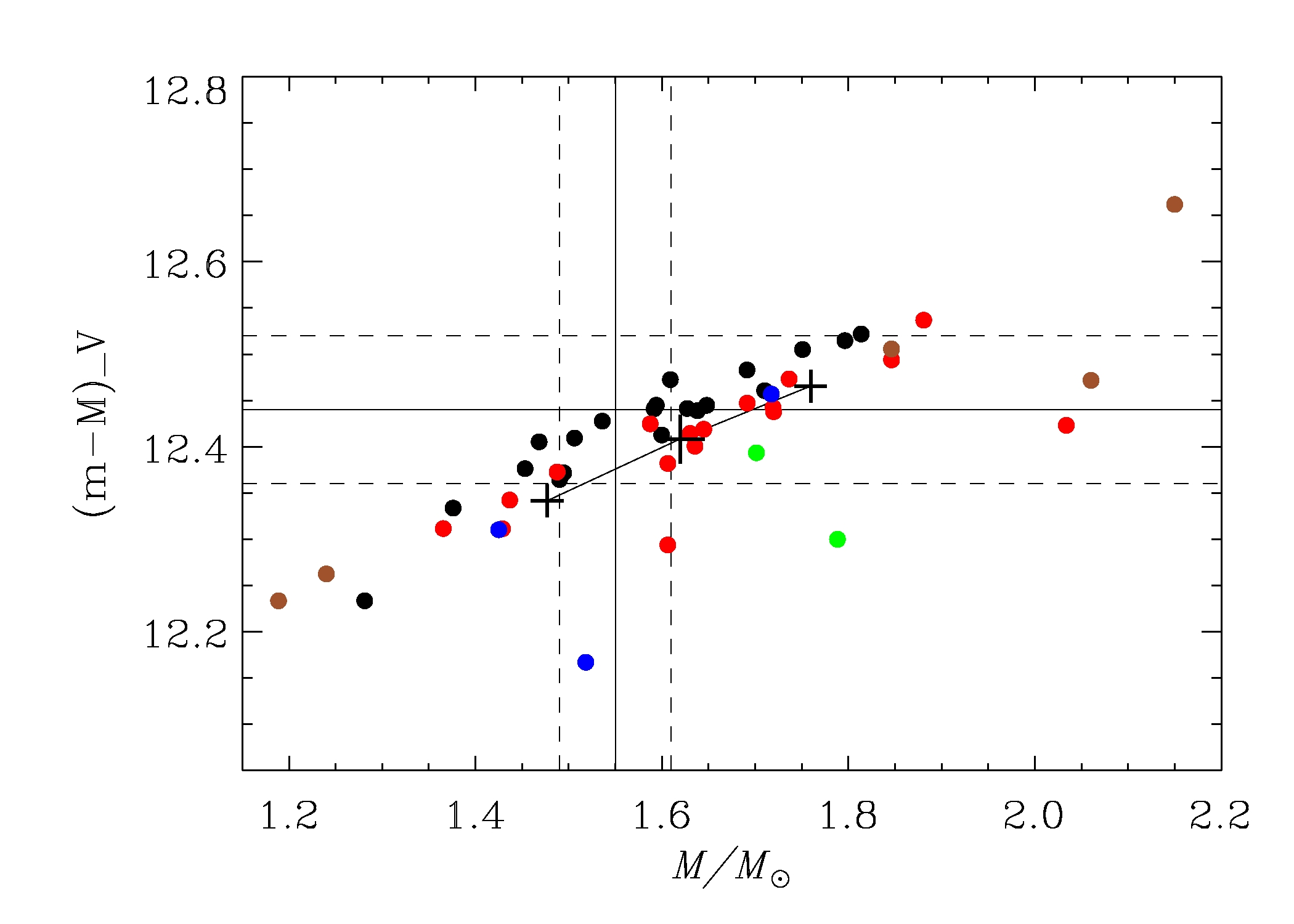}
%
% If no graphics program available, insert a blank space i.e. use
%\picplace{5cm}{2cm} % Give the correct figure height and width in cm
%
\caption{As Fig. \ref{fig:1} but using bolometric corrections and colour-temperature relations from \cite{Casagrande13} and the correction to the scaling relations from \cite{Mosser13}. The relative correction between RGB and RC stars determined by \cite{Miglio12} is still applied to the RC stars.}
\label{fig:3}       % Give a unique label
\end{figure}

Fig. \ref{fig:2} shows the situation when all measurements are made using the bolometric corrections and colour-temperature calibrations of \cite{Casagrande13} based on MARCS models (while still in preparation, those calibrations were already used in several investigations and some details are available in \cite{VandenBerg13}). We assumed $E(B-V)=0.15$ and $E(V-Ks)/E(B-V)=2.72$ since this results in $T_{\rm{eff}}$ for the red clump stars in agreement with spectroscopic measurements \citep{Bragaglia01}.
This shifts the asteroseismic measurements such that there is an overlap region with the binary results. However, to shift the mean asteroseismic mass close to the binary result, a correction to the scaling relations was still needed. We have therefore also applied the correction to the scaling relations suggested by \cite{White11} in Fig. \ref{fig:2}. That way, we can obtain a reasonable agreement! However, as shown in Fig. \ref{fig:3}, a similar agreement can be obtained by using instead the correction suggested by \cite{Mosser13}. In principle, the correct correction might be expected to decrease the scatter among the measured masses and distance moduli, but unfortunately the different corrections are too similar and random uncertainties too large to allow such a test (the small plus signs attached to the mean value in Fig. \ref{fig:3} shows the effect on the cluster mean when $\nu _{\rm{max}}$ is changed by $\pm 1-\sigma$ for all stars). Thus, we are left in a situation where indications are that the asteroseismic scaling relations overestimate masses of giant stars, but we cannot distinguish between different suggested corrections, because they both work equally well, at least in this situation.

The correction suggested by \cite{White11} depends on $T_{\rm{eff}}$ and similar corrections have been calculated by \cite{Miglio13}, where they also extend to the helium-burning red clump phase of evolution, showing a different correction to red clump stars compared to stars in the red giant branch phase (see the details in Fig. 5 of \citealt{White11} and Fig. 2 of \citealt{Miglio13}). Note especially that for the case of the open cluster NGC6811, where the giants are younger and hotter, the correction predicted by \cite{Miglio13} becomes small while for the red clump stars of the even younger clusters NGC6866 and NGC6633 this correction becomes large, but in the opposite direction! On the other hand, the correction suggested by \cite{Mosser13} does not depend on $T_{\rm{eff}}$. Therefore, by extending our investigation to other clusters of different ages, and therefore different masses and $T_{\rm{eff}}$'s for the giant stars, it should be possible to find a suitable form of a correction. We are already in the process of identifying, observing and analysing detached eclipsing systems and oscillating giant stars in the open clusters NGC6791, NGC6811, NGC6866, and NGC6633.

There are, however, some additional difficulties in this procedure. One issue we have noticed is that the $T_{\rm{eff}}$ of the red clump stars measured spectroscopically for NGC6819 \citep{Bragaglia01} and NGC6811 \citep{Joanna13} are higher than the corresponding $T_{\rm{eff}}$ calculated by \cite{Miglio13} for the corresponding asteroseismic masses. As shown in the contribution by A. Miglio in this volume, different models predict different $T_{\rm{eff}}$ for the giant stars. This introduces an additional challenge when wanting to apply a correction to the asteroseismic scaling relations based on model $T_{\rm{eff}}$'s.

It is also not currently known whether a correction that depends directly on the metallicity should be made (to either $\Delta \nu$ or $\nu_{\rm{max}}$, or perhaps both?). Unfortunately, there are no metal-poor clusters in the sample of open cluster in the {\it Kepler} and CoRoT fields, so no such tests can presently be done. In order to check for possible metallicity effects, we are instead working on potentially metal-poor detached eclipsing binaries in the {\it Kepler} field which contain a giant star showing solar-like oscillations. We will analyse these along the lines of \cite{Frandsen13} and compare directly to the asteroseismic signal of the giant component of the binary. Our sample contains three systems that have [Fe/H] $\approx -0.4$ according to the {\it Kepler} input catalogue.

\section{Identification of stars that evolved through non-standard evolution}
\label{sec:3}
% Always give a unique label
% and use \ref{<label>} for cross-references
% and \cite{<label>} for bibliographic references
% use \sectionmark{}
% to alter or adjust the section heading in the running head

The red giant phase of evolution is so short-lived that the difference in mass among the giants in a cluster is much smaller than our asteroseismic measurement uncertainty. Therefore, it makes sense to use the ensemble mean mass as the most precise measure of the mass of an RGB stars in a cluster. However, if there are giants in the sample that did not evolve as a single star, they are likely to have a different mass and should be excluded when calculating a mean mass from the ensemble.

An asteroseismic study by \cite{Corsaro12} has already identified such stars in NGC6819 by investigating the period spacing of dipole modes caused by gravity modes in the oscillation spectra, $\Delta P_{\rm{obs}}$. By comparing their Fig. 8 diagram of $\Delta \nu$ versus $\Delta P_{\rm{obs}}$ to the corresponding diagram in Fig. 4 of \cite{Stello13} calculated using models, one sees that for red clump stars, the differences in $\Delta \nu$ are caused by a combination of differences in mass {\it and} evolution. Returning to Fig. \ref{fig:3} we find a red clump star with a mass of $2.05 M_\odot$. It turns out that this star, KIC5023953, sits in the expected location for normal red clump stars of NGC6819 in Fig. 8 of \cite{Corsaro12}, despite our evidence of a larger mass. But since the red clump sequence in that diagram is also affected by evolution during the helium-burning phase, the star is consistent with being an over-massive star well into the helium-burning phase. The fact that the star is a binary \citep{Stello11, Hole09} supports this interpretation, since the higher mass could have originated from mass-transfer. This example shows that it is wise to combine information from several diagrams before drawing definitive conclusions about specific stars. Any star not lying along the scatter line of the majority of stars in a diagram like Fig. \ref{fig:3} should be investigated for signs of non-standard evolution.

\section{Conclusions}

We have shown indications that the asteroseismic scaling relations for solar-like oscillators (Eq. \ref{eq:01}) will overestimate the mass of a star unless some kind of correction is applied. At present we are unable to distinguish between different corrections suggested in the literature, since they work equally well. An ongoing expanded investigation employing eclipsing binaries in more open clusters and in the field will allow more firm conclusions to be made.

\begin{acknowledgement}
K. Brogaard acknowledges funding from the Villum Foundation.
\end{acknowledgement}
%

%\input{referenc}
% BibTeX users please use
%\bibliographystyle{../aa.bst}
%\bibliography{brogaard_sesto}

\end{document}